\documentclass[nonacm, sigconf, anonymous=false]{acmart}

\usepackage{amsmath}
\usepackage[linesnumbered,ruled]{algorithm2e}
\usepackage{bm}  

\SetKwInput{KwInput}{Input}                
\SetKwInput{KwOutput}{Output}              
\SetKwInput{KwParticipant}{Participants}              
\SetKwInput{kwConstraint}{Constraint}      
\SetKwBlock{KwTA}{$P_2$}{end}
\SetKwBlock{KwCSP}{$P_3$}{end}
\SetKwBlock{KwCK}{$C_K$}{end}
\SetKwBlock{KwDH}{Data Holders}{end}
\SetKwFunction{MUL}{Mul}
\SetKwFunction{SDS}{SecretDataSharing}
\SetKwFunction{SFE}{SecureFitnessEvaluation}
\SetKwFunction{FFedPCAPredict}{FedPCA.predict}
\SetKwFunction{FFedMPCAPredict}{FedMPCA.predict}
\SetKwProg{Fn}{Function}{:}{End}

\setcopyright{none}
\renewcommand\footnotetextcopyrightpermission[1]{}

\AtBeginDocument{%
  \providecommand\BibTeX{{%
    \normalfont B\kern-0.5em{\scshape i\kern-0.25em b}\kern-0.8em\TeX}}}

\begin{document}

\title{Towards Vertical Privacy-Preserving Symbolic Regression via Secure Multiparty Computation}
\author{Du Nguyen Duy}
\email{du.nguyen.duy@scch.at}
\affiliation{%
  \institution{Software Competence Center Hagenberg}
  \streetaddress{Softwarepark 32a}
  \city{Hagenberg}
  \country{Austria}
  \postcode{4232}
}

\author{Michael Affenzeller}
\email{michael.affenzeller@heuristiclab.com}
\affiliation{%
  \institution{University of Applied Sciences Upper Austria}
  \streetaddress{Softwarepark 11}
  \city{Hagenberg}
  \country{Austria}}
  \postcode{4232}

\author{Ramin Nikzad-Langerodi}
\email{ramin.nikzad-langerodi@scch.at}
\affiliation{%
  \institution{Software Competence Center Hagenberg}
  \streetaddress{Softwarepark 32a}
  \city{Hagenberg}
  \country{Austria}
  \postcode{4232}
}

\renewcommand{\shortauthors}{Du, et al.}

\begin{abstract}
Symbolic Regression is a powerful data-driven technique that searches for mathematical expressions that explain the relationship between input variables and a target of interest. Due to its efficiency and flexibility, Genetic Programming can be seen as the standard search technique for Symbolic Regression. However, the conventional Genetic Programming algorithm requires storing all data in a central location, which is not always feasible due to growing concerns about data privacy and security. While privacy-preserving research has advanced recently and might offer a solution to this problem, their application to Symbolic Regression remains largely unexplored. Furthermore, the existing work only focuses on the horizontally partitioned setting, whereas the vertically partitioned setting, another popular scenario, has yet to be investigated. Herein, we propose an approach that employs a privacy-preserving technique called Secure Multiparty Computation to enable parties to jointly build Symbolic Regression models in the vertical scenario without revealing private data. Preliminary experimental results indicate that our proposed method delivers comparable performance to the centralized solution while safeguarding data privacy.
\end{abstract}





\maketitle

\section{Introduction}

Symbolic Regression (SR) is a machine learning technique that seeks to find the underlying mathematical relationship between the input and output variables without imposing any preconceived notions about the structure of the model. Compared to other commonly used data-driven methods, such as Support Vector Machines or Deep Neural Networks, it has the advantage of producing models that are easy to interpret and understand since the resulting expression can be represented in a form that is familiar to humans \cite{affenzeller2014}. This can be particularly useful in fields such as physics, engineering, and finance, where the ability to understand and analyze the hidden equations is critical.

Traditional methods of SR involve using evolutionary algorithms, particularly Genetic Programming (GP) \cite{koza1994}. In GP-based SR (GPSR), a population of candidate expressions is evolved using genetic operators, such as selection, crossover, and mutation, to improve a fitness function (e.g., Mean Squared Error) that evaluates the quality of the expressions. While GPSR can be effective, growing concerns about data privacy and ownership might limit its applicability to real-world scenarios. More specifically, during model training, the whole dataset is needed to evaluate the quality of solutions. Hence, the existing GP methods often assume that data is stored in one place (i.e., a server or a cluster) and can be easily accessible. However, this assumption is very challenging, if not impossible, to satisfy when the data is distributed among different parties who might not share the same interests. In such situations, even though data from all parties must be analyzed simultaneously to achieve optimal performance, data centralization faces many challenges.

On the one hand, data owners might be unwilling to share private data because it is valuable or incorporates sensitive information. On the other hand, nowadays, there are laws and regulations that restrict organizations from freely exchanging data \cite{yang2019}. Overall, data privacy and security have become a prevalent topic not only for GPSR but for all other data-driven techniques. During the last few years, many frameworks have been proposed for training models while upholding data privacy and ownership, such as Federated Learning (FL) \cite{konevcny2016}, Secure Multi-party Computation (MPC) \cite{mohassel2017}, Homomorphic Encryption (HE) \cite{gentry2009}, and Differential Privacy (DP) \cite{dwork2014}. Based on these studies, numerous privacy-preserving machine learning algorithms have been developed. The existing works include linear regression \cite{gascon2016}, tree-based models \cite{liu2020}, neural network architectures \cite{wagh2019}, and latent variables-based models \cite{du2022}. Nevertheless, relatively limited attention has been paid to their adoption in GPSR, considering its wide-ranging potential.

As mentioned previously, the challenge in performing GPSR in a privacy-preserving manner is to evaluate the fitness value on data scattered across different organizations. There are two common data distribution schemas: horizontal setting and vertical setting. Horizontal setting refers to a scenario where  different parties possess datasets that share the same variable space but have minimal overlap in the sample space. This partitioning of data is a common approach, particularly in the cross-device setting, where various users aim to enhance their model performance for the same task. In contrast, in a vertical setting, the local datasets share the same sample space but differ in the feature space. This scheme is especially common in cooperation among different enterprises across a value chain. While an FL-based GP framework for training SR on horizontally distributed data has been proposed recently \cite{dong2022}, an approach for performing secure GPSR in the vertical scenario is still missing.

To this end, we propose an MPC-based framework called Privacy Preserving Symbolic Regression (PPSR) for training symbolic regression models on vertically distributed data. To the best of our knowledge, this is the first study that addresses this problem. In the context of the preliminary experiment, PPSR yields comparable performance to the centralized GPSR approach on the simulated datasets.

\section{Background}

\subsection{Symbolic Regression}

\begin{figure}[!ht]
    \centering
    \includegraphics[width=0.25\textwidth]{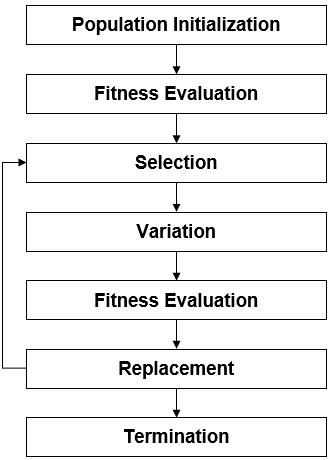}
    \caption{An illustration of the GPSR workflow.}
    \label{fig:gpsr}
\end{figure}

\begin{figure}[!ht]
    \centering
    \includegraphics[width=0.35\textwidth]{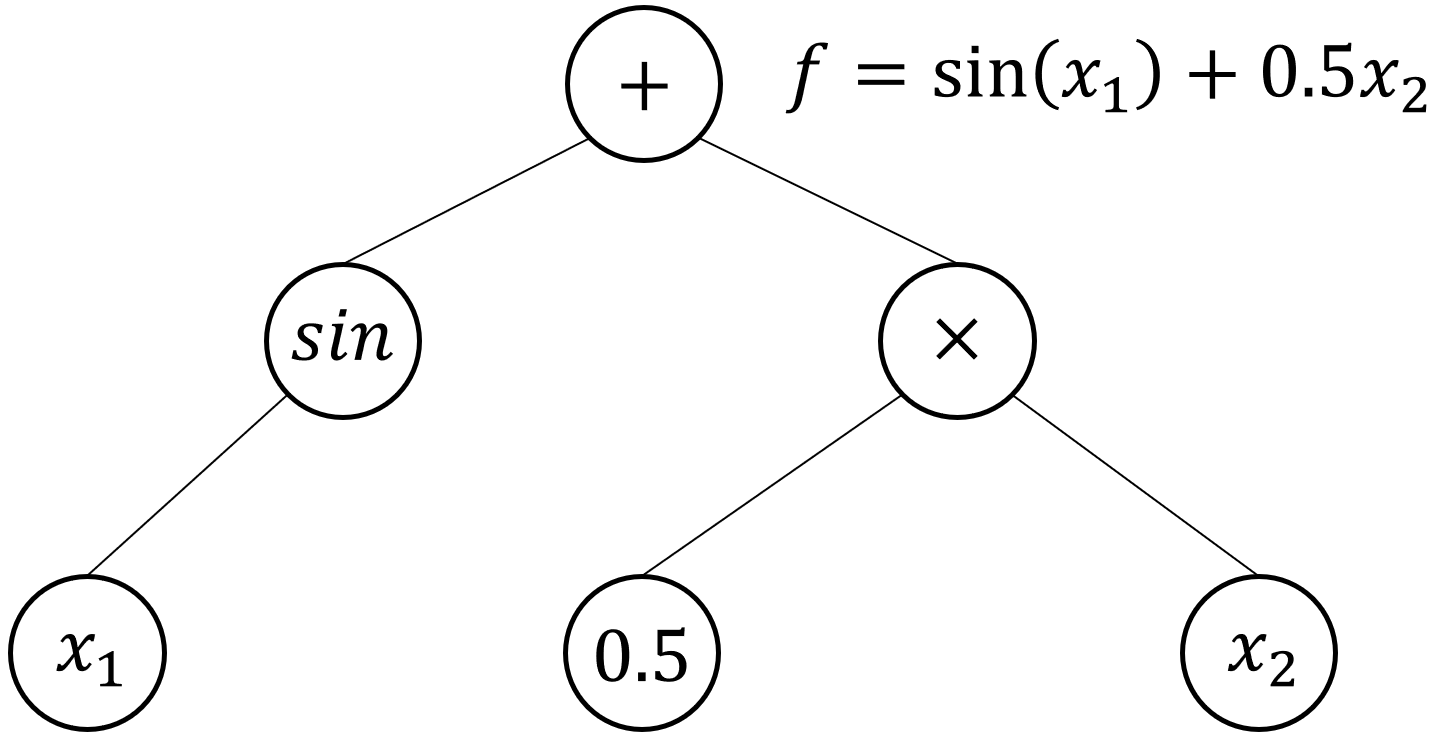}
    \caption{An illustration of a tree structure.}
    \label{fig:tree}
\end{figure}

As mentioned previously, Symbolic Regression (SR) is a form of regression analysis that aims to find a mathematical expression (a symbolic model) that best describes the relationship between a set of independent variables and a dependent variable of a given dataset. When Genetic Programming (GP) is used for conducting SR, the overall process (depicted in \textbf{Figure \ref{fig:gpsr}}) can be described step-by-step as follows:

\begin{itemize}
    \item \textbf{Step 1: Population Initialization}. The first step is to initialize a population of candidate solutions. Each candidate solution is represented as a tree structure (see \textbf{Figure \ref{fig:tree}}), where the nodes represent mathematical operators ($+$, $-$, $\times$, /, $sin(\cdot)$, $cos(\cdot)$, ...), and the leaves represent variables, constants or coefficients.

    \item \textbf{Step 2: Fitness Evaluation}. Each candidate solution is evaluated by computing its fitness on the given dataset. This is done by comparing the candidate solution's predicted values with the dependent variables' actual values.

    \item \textbf{Step 3: Selection}. The fittest candidate solutions are selected for reproduction. Selection is typically made using a fitness-based approach, such as tournament selection or roulette wheel selection.

    \item \textbf{Step 4: Variation}. The selected candidate solutions are modified through a combination of crossover and mutation operators. In the standard case, crossover involves swapping subtrees between two candidate solutions, while mutation involves randomly changing a subtree in a candidate solution.

    \item \textbf{Step 5: Fitness Evaluation}. Evaluate the fitness of the new candidate solutions by calculating the error between the predicted output of the solution and the actual output of the training data.

    \item \textbf{Step 6: Replacement}. Replace the old population with the new population of candidate solutions. The fittest solutions from the previous generation may be kept, while others may be replaced by the new solutions.

    \item \textbf{Step 7: Termination}. Repeat steps 3 to 6 until a termination criterion is met. This criterion can be a maximum number of generations, a minimum fitness threshold, or a maximum amount of computation time. The final generation of candidate solutions is typically sorted by fitness, and the fittest candidate solution is selected as the solution to the symbolic regression problem. 
\end{itemize}

During the entire process, Fitness Evaluation is the only module where the algorithm needs to access the whole dataset. Thus, how to perform it while protecting data privacy is the key to enabling the construction of a GP-based SR model in a privacy-preserving manner.

Suppose the given data is expressed as:
\begin{equation}
    D = \{d_i \ | \ d_i = (x_{i1}, x_{i2},..., x_{in}, y_i), i = 1, 2,..., m\}
\end{equation}
where $n$ represents the number of input variables, $m$ is the number of sample data, $x_{ik}$ means the $k^{th}$ variable value of the $i^{th}$ sample data, and $y_i$ denotes the corresponding target.

The fitness function $F(\cdot)$ is typically a measure of how well a candidate solution $f(\cdot)$ fits the training data $D$, and it can take various forms depending on the specific problem and the goals of the analysis. A widely used fitness function is Mean Square Error ($MSE$), which is defined as follows:
\begin{equation}
    F = \frac{1}{m}\sum_{i=1}^{m}(y_i - \hat{y}_i)^2
\end{equation}
where $\hat{y}_i = f(x_{i1}, x_{i2}, ..., x_{in})$ is the predicted value for the $i^{th}$ observation using the candidate solution.

Other often used functions are Root Mean Squared Error ($RMSE$) and R-Squared ($R^2$). However, regardless of the choice of the fitness function, the same problem arises when $D$ is jointly owned by different clients who do not want to expose private data.

\subsection{Horizontal Federated Symbolic Regression}



\begin{figure}[!ht]
    \centering
    \includegraphics[width=0.45\textwidth]{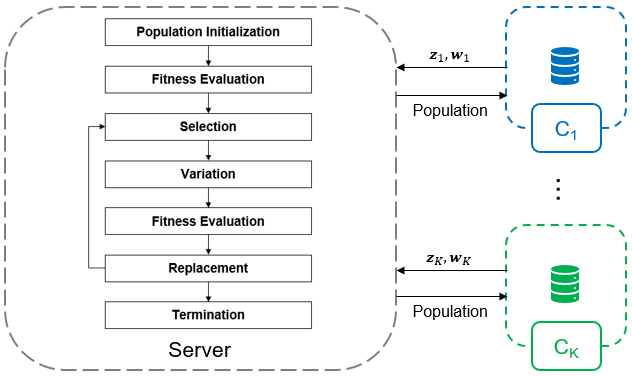}
    \caption{Federated Genetic Programming framework.}
    \label{fig:fedgp}
\end{figure}

In \cite{dong2022}, the authors proposed a horizontal federated GP framework that enables the construction of GPSR models without disclosing local data from data holders. The proposed framework consists of multiple clients labeled as $C_1$ to $C_K$ and a central server, as shown in \textbf{Figure \ref{fig:fedgp}}. Initially, the server generates a population and broadcasts it to all clients. Then, the clients calculate local fitness for all solutions (denoted as a vector $\mathbf{z}_j$) in parallel using private data and then send local fitness along with the local weight (denoted as $w_j$) back to the server. The local weight indicates the client's relative impact on the global model and is proportional to the amount of data contributed. The server uses a non-parametric clustering technique called Mean Shift Aggregation \cite{dong2022} to estimate global fitness values. Finally, the server performs genetic operations to create a new population and then repeats the process until the stopping criteria are met.

The federated architecture employed here helps to protect privacy as it allows local clients to keep their training data confidential while transmitting only the fitness information to the server. However, since it assumes clients can calculate the fitness value independently, this framework is only feasible in the horizontal setting when the local datasets share the same feature space. In contrast, in a vertical scenario, when the generated expressions involve variables belonging to different clients, no single client can estimate the fitness alone because it cannot access all the involved variables. Therefore, the key distinction between our method and this work is our ability to handle vertically distributed data, which is achieved by leveraging a privacy-preserving technique called Secure Multiparty Computation.

\subsection{Secure Multiparty Computation}\label{mpc}

Secure Multiparty Computation (MPC) is a cryptographic approach that enables computation on data from different clients, such that the participants gain no additional information about each others’ inputs, except what can be learned from the public output of the algorithm. The underlying mechanics depend on the protocol being applied. In this work, we use the 3-party secure computation protocol (3PC), which is a framework extensively employed in various privacy-preserving machine learning systems \cite{demmler2015}\cite{wagh2019}. 

\begin{figure}[!ht]
    \centering
    \includegraphics[width=0.35\textwidth]{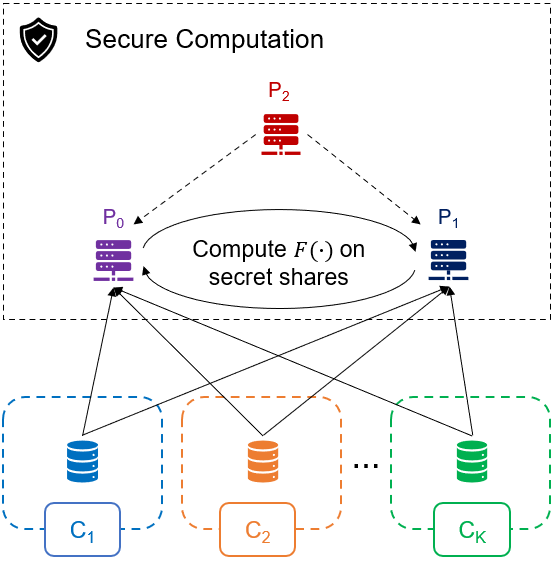}
    \caption{An illustration of the 3PC scheme.}
    \label{fig:3pc_scheme}
\end{figure}

In the 3PC scheme, the three parties involved are denoted as $P_0$, $P_1$, and $P_2$ (see \textbf{Figure \ref{fig:3pc_scheme}}). All operations are performed in the ring $\mathbb{Z}_{2^L}$ (integers modulo $2^L$), where each number is represented as an $L$-bit integer. To securely evaluate a function $F(\cdot)$, the data owners first share their private values with $P_0$ and $P_1$ using a secret sharing scheme. Then $P_0$ and $P_1$ will proceed to evaluate $F(\cdot)$ over the secret shares with the support of $P_2$. The output will be similarly secret-shared, meaning $P_0$ and $P_1$ each own just a share of the result. Only when the shares are recombined the real result is revealed. 

The sharing of a value $x$ is denoted by $\langle x \rangle = \{\langle x\rangle_0, \langle x\rangle_1\}$, where $\langle x\rangle_i \in \mathbb{Z}_{2^L}$ indicates $P_i$'s share of $x$. To share a secret $x$, the secret holder picks a random number $r$ from $\mathbb{Z}_{2^L}$, assigns $\langle x\rangle_0 := r$ and $\langle x\rangle_1:=x - r$, and then sends $\langle x\rangle_0$ and $\langle x\rangle_1$ and to $P_0$ and $P_1$ respectively. Because of the way the shares are created, from $P_i$'s perspective, they are merely random numbers. However, when they are both sent to some party $P_i$, $P_i$ can reconstruct $x$ immediately by summing up the two shares, i.e., $x = \langle x\rangle_0 + \langle x\rangle_1$.

Due to their homomorphic properties, the secret shares can be utilized to implement secure computations. Addition and multiplication operations can be easily performed on the shared values as follows:
\begin{itemize}
    \item Addition/Subtraction of a secret shared value $x$ to a public value (values known to all parties) $a$: $P_0$ maintains $\langle x \rangle_0 \pm a$, and $P_1$ keeps $\langle x \rangle_1$.
    \item Multiplication of a secret shared value $x$ with a public value $a$: The two parties multiply their shares with $a$ respectively, i.e., $P_0$ gets $\langle x \rangle_0 \cdot a$, $P_1$ gets $\langle x \rangle_1 \cdot a$.
    \item Addition/Subtraction of two secret shared values $x$ and $y$: The two parties add/subtract their shares respectively, i.e., $P_0$ gets $\langle x \rangle_0 \pm \langle y \rangle_0$, and $P_1$ gets $\langle x \rangle_1 \pm \langle y \rangle_1$.
    \item Multiplication of two secret shared values $x$ and $y$ is implemented using random Beaver triples \cite{beaver1992}, ($a$, $b$, $c$) with $c = ab$, that $P_2$ provides. The details of this operation are listed in \textbf{Algorithm \ref{alg:ss_mul}}. It is straightforward to prove that $\langle xy \rangle_0 + \langle xy \rangle_1 = xy$.
\end{itemize}

\begin{algorithm}
\DontPrintSemicolon
  \KwParticipant{$P_0$, $P_1$, and $P_2$}
  \KwInput{$P_i$ holds \{$\langle x \rangle_i$, $\langle y \rangle_i$\}, $i \in \{0, 1\}$}
  \KwOutput{$P_i$ gets $\langle xy \rangle_i$, $i \in \{0, 1\}$}
  \Fn{\MUL{$\langle x \rangle$, $\langle y \rangle$}}{
         \KwTA(do:){
            Generate a Beaver triple $(a, b, c)$ such that $c = ab$\\
            Generate for $i \in \{0, 1\}$, $\langle a \rangle_i , \langle b \rangle_i, \langle c \rangle_i$ and send to $P_i$ 
         }
       \For{$i \in \{0,1\}$, $P_i$}{
            $\langle \epsilon \rangle_i = \langle x \rangle_i - \langle a \rangle_i$\\
            $\langle \delta \rangle_i = \langle y \rangle_i - \langle b \rangle_i$\\
            Reconstruct $\epsilon$ and $\delta$ by exchanging shares. $\epsilon$ and $\delta$ are now public values.\\
            Compute $\langle xy \rangle_i = \langle c \rangle_i + \epsilon\langle b \rangle_i + \langle a \rangle_i\delta + i\epsilon\delta$
       }
  }
\caption{Secret Sharing Multiplication}
\label{alg:ss_mul}
\end{algorithm}

The combination of addition and multiplication operations can be used to calculate linear and polynomial functions. For more complex functions that cannot be efficiently computed using only these operations, standard approximations can be utilized.

In our implementation, we leverage a Python package called CRYPTEN \cite{knott2021} for performing secure computations. This library employs limit approximation to compute exponentials, Householder iterations to compute logarithms, Newton-Raphson iterations to compute reciprocals, and the repeated-squaring method to compute sine and cosine functions. As 3PC is performed on the integer ring $\mathbb{Z}_{2^L}$, when the inputs are in floating-point format, they must be converted into fixed-point. CRYPTEN handles this conversion by using fixed-point encoding \cite{mohassel2017}. More specifically, to obtain an integer value $x$ from a floating-point value $x_\mathbb{R}$, it multiplies $x_\mathbb{R}$ with a large scaling factor $2^B$ and rounds the result to the nearest integer $x = \lfloor 2^Bx_\mathbb{R} \rceil$. And to decode an integer value, $x$, it calculates $x_\mathbb{R} \approx x/2^B$. This method is naturally suitable for addition and subtraction operations. However, considering the multiplication of two numbers $x_\mathbb{R}$ and $y_\mathbb{R}$, we get $\lfloor 2^Bx_\mathbb{R} \rceil \lfloor 2^Bx_\mathbb{R} \rceil \approx 2^B\lfloor 2^Bx_\mathbb{R}y_\mathbb{R} \rceil$. Therefore, after every multiplication, the output will be scaled down by a factor of $2^B$.

\section{Vertical Privacy-Preserving Symbolic Regression}

\subsection{Preliminaries}

In this study, we concentrate on the situation where $K$ clients $C_1, C_2, ..., C_K$ want to train an SR model on the joint data vertically partitioned among them. Suppose $C_j$ secretly owns $\mathbf{X}_j \in \mathbb{Z}^{m \times n_j}_{2^L}$ where $m$ represents the number of observations, and $n_j$ represents the number of private variables. In this scenario, only $C_K$ has access to the target variable $\mathbf{y} \in \mathbb{Z}^{m \times 1}_{2^L}$. The joint dataset can be denoted as $D = \{\mathbf{X}, \mathbf{y}\}$ where $\mathbf{X} = [\mathbf{X}_1 | \mathbf{X}_2 | ... | \mathbf{X}_K] \in \mathbb{Z}^{m \times n}_{2^L}$ with $n = \sum_{j = 1}^{K} n_j$.

Even though we do not explicitly mention the conversion between real-valued numbers and integer numbers in this section, it can be assumed to occur and solved using the fixed-point encoding protocol explained previously.

\subsection{Methodology}

The proposed framework requires four parties (servers) which are labelled as $P_0$, $P_1$, $P_2$, and $P_3$. While any client can assume the role of $P_3$, $P_2$ must be a third party trusted by all clients. When there are only two clients, they themselves can operate as $P_0$ and $P_1$. If more clients are involved, two of them can be promoted as $P_0$ and $P_1$. The overall workflow of PPSR includes two phases: Secret Data Sharing and Model Training.

\begin{figure*}[!ht]
    \centering
    \includegraphics[width=0.75\textwidth]{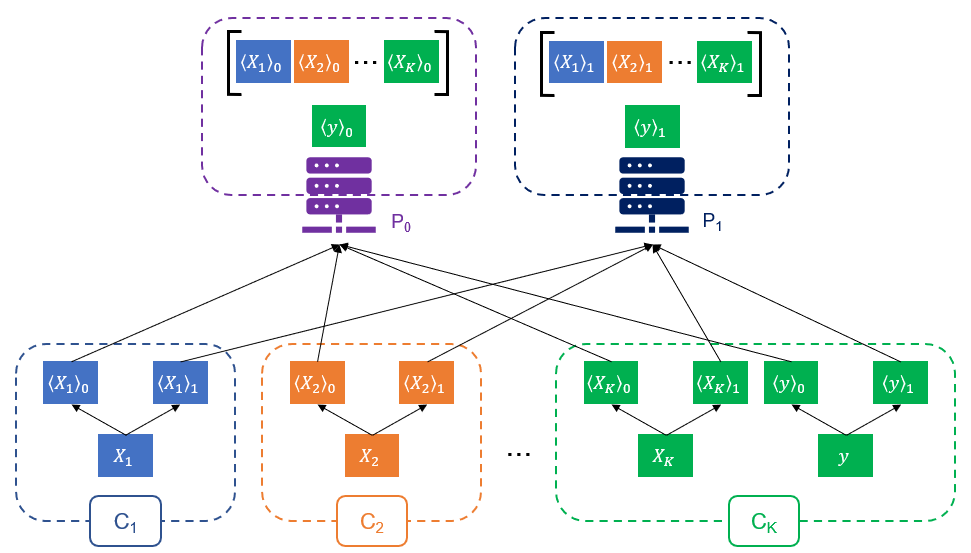}
    \caption{An illustration of the Secret Data Sharing phase.}
    \label{fig:data_sharing}
\end{figure*}

The phase of Secret Data Sharing involves all clients uploading their private data to $P_0$ and $P_1$ in a secret-shared manner, as depicted in \textbf{Figure \ref{fig:data_sharing}}. In \textbf{Section \ref{mpc}}, the sharing of a single value has been explained. To securely share a matrix $\mathbf{X}_j$ or a vector $\mathbf{y}$ (only $C_K$), clients simply apply secret sharing element-wise. The details of this process are specified in \textbf{Algorithm \ref{alg:secret_data_sharing}}. As a result, each $P_i$ holds a share of $\mathbf{X}$ and $\mathbf{y}$, referred to as $\langle \mathbf{X} \rangle_i$ and $\langle \mathbf{y} \rangle_i$ respectively, such that $\sum_{i=0}^{1} \langle \mathbf{X} \rangle_i = X$ and $\sum_{i=0}^{1} \langle \mathbf{y} \rangle_i = \mathbf{y}$. It is strictly required that $P_0$ and $P_1$ are trustworthy and do not collude; otherwise, they could exchange the shares and gain access to the actual dataset.

Once the data is shared with the computing parties ($P_0$ and $P_1$), the Model Training phase commences at $P_3$. The process involves Population Initialization, Fitness Evaluation, Selection, Variation, Replacement, and Termination, which is similar to the standard GP process. However, the unique aspect of PPSR is in the Fitness Evaluation step where fitness values are computed over the secret shares held by $P_0$ and $P_1$ rather than using the raw data.

\begin{figure}[!ht]
    \centering
    \includegraphics[width=0.40\textwidth]{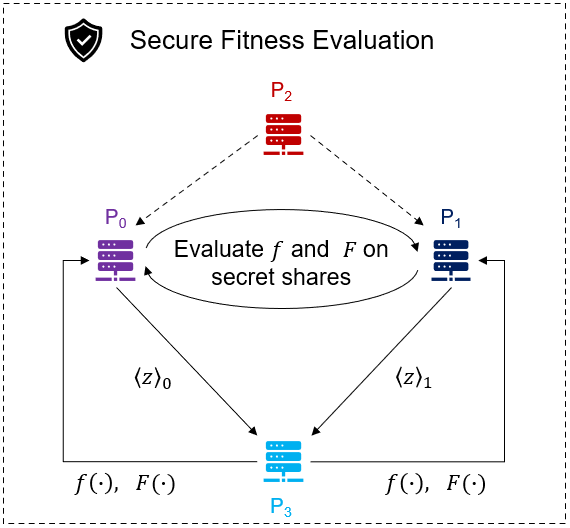}
    \caption{An illustration of the Secure Fitness Evaluation module.}
    \label{fig:secure_fitness_evaluation}
\end{figure}

In order to assess the quality of a candidate expression $f(\cdot)$, $P_3$ will forward both the function and the fitness function $F(\cdot)$ (e.g., $MSE$, $RMSE$, $R^2$) to $P_0$ and $P_1$. Then these two parties will estimate their shares of the predicted values ($\langle \hat{\mathbf{y}} \rangle_i$) by iteratively evaluating the function $f(\cdot)$ on each row of $\langle \mathbf{X} \rangle_i$. As $\langle \hat{\mathbf{y}} \rangle_i$ is an output of computations over secret shares, it is simply a vector of random numbers from $P_i$'s viewpoint. However, by definition, adding $\langle \hat{\mathbf{y}} \rangle_0$ to $\langle \hat{\mathbf{y}} \rangle_1$ reveals the actual predicted values $\hat{y}$. $P_0$ and $P_1$ then calculate their shares of the fitness value $\langle z \rangle_i$ by applying the fitness function $F(\cdot)$ on $\langle \mathbf{y} \rangle_i$ and $\langle \hat{\mathbf{y}} \rangle_i$, and send $\langle z \rangle_i$ back to $P_3$ (depicted in \textbf{Figure \ref{fig:secure_fitness_evaluation}}). Finally, $P_3$ recombines the shares to obtain the true fitness value $z$. The algorithmic representation of the secure fitness evaluation protocol is displayed in \textbf{Algorithm \ref{alg:secure_fitness_evaluation}}.

\begin{algorithm}
\DontPrintSemicolon
  \KwParticipant{$C_j$ for $j \in \{1, 2,..., K\}$, $P_0$, and $P_1$}
  \KwInput{$C_j$ holds $\mathbf{X}_j \in \mathbb{Z}^{m \times n_j}_{2^L}$ for $j \in \{1, 2,..., K\}$, $C_K$ holds $\mathbf{y} \in \mathbb{Z}^{m \times 1}_{2^L}$}
  \KwOutput{For $i \in \{0, 1\}$, $P_i$ gets their share of the training data $\langle D \rangle_i$}
  \Fn{\SDS{}}{
    \For{$j \in \{1, 2,..., K\}$, $C_j$}{
        Generate a random matrix $\mathbf{R} \in \mathbb{Z}_{2^L}^{m \times n_j}$\\
        Calculate $\langle \mathbf{X}_j \rangle_0 = \mathbf{R}$ and $\langle \mathbf{X}_j \rangle_1 = \mathbf{X}_j - \mathbf{R}$\\
        Transfer $\langle \mathbf{X}_j \rangle_0$ and $\langle \mathbf{X}_j \rangle_1$ to $P_0$ and $P_1$ respectively
    }
    \KwCK(do:){
        Generate a random vector $\mathbf{w} \in \mathbb{Z}_{2^L}^{m \times 1}$\\
        Calculate $\langle \mathbf{y} \rangle_0 = \mathbf{w}$ and $\langle \mathbf{y} \rangle_1 = \mathbf{y} - \mathbf{w}$\\
        Transfer $\langle \mathbf{y} \rangle_0$ and $\langle \mathbf{y} \rangle_1$ to $P_0$ and $P_1$ respectively
    }
    \For{$i \in \{0, 1\}$, $P_i$}{
        Compile their share of the training dataset: $\langle D \rangle_i = \{ \langle \mathbf{X} \rangle_i, \langle \mathbf{y} \rangle_i \}$, with $\langle \mathbf{y} \rangle_i \in \mathbb{Z}^{m \times 1}_{2^L}$ and $\langle \mathbf{X} \rangle_i = [\langle \mathbf{X}_1 \rangle_i | \langle \mathbf{X}_2 \rangle_i | ... | \langle \mathbf{X}_K \rangle_i] \in \mathbb{Z}^{m \times n}_{2^L}$
    }
  }
\caption{Secret Data Sharing}
\label{alg:secret_data_sharing}
\end{algorithm}

\begin{algorithm}
\DontPrintSemicolon
  \KwParticipant{$P_0$, $P_1$, $P_2$, and $P_3$}
  \KwInput{$P_i$ holds $\langle D \rangle_i$ for $i \in \{0, 1\}$, $P_3$ holds the candidate expression $f(\cdot)$ and the fitness function $F(\cdot)$}
  \KwOutput{$P_3$ gets the fitness value $z$ of $f(\cdot)$}
  \Fn{\SFE{}}{
    \KwCSP(do:){
        Forward the candidate expression $f(\cdot)$ and the fitness function $F(\cdot)$ to $P_0$ and $P_1$
    }
    \For{$i \in \{0,1\}$, $P_i$}{
        Compute its shares of predictions $\langle \hat{\mathbf{y}} \rangle_i$ locally by evaluating $f(\cdot)$ on each row of $\langle \mathbf{X} \rangle_i$\\
        Compute its share of the fitness value $\langle z \rangle_i$ locally by evaluating $F(\cdot)$ on $\langle \mathbf{y} \rangle_i$ and $\langle \hat{\mathbf{y}} \rangle_i$\\
        Send $\langle z \rangle_i$ to $P_3$
    }
    *The assistance of $P_2$ is required whenever multiplication is involved.\\
    \KwCSP(do:){
            Combine the secret shares and get the real fitness value: $z = \langle z \rangle_0 + \langle z \rangle_1$
    }
  }
\caption{Secure Fitness Evaluation}
\label{alg:secure_fitness_evaluation}
\end{algorithm}

\subsection{Security Justification}\label{security_justification}

In this section, we justify the security of PPSR. We consider the semi-honest scenario, which covers a wide range of secure machine learning use cases. Under this setting, each party will strictly comply with the predefined protocol but might attempt to gain as much information as possible from their perspective \cite{evans2018}. In addition, we also assume that there is no collusion between parties and clients.

In our proposed framework, throughout the entire process, the real datasets stay securely in the client's database. Even though the secret shares are transferred to computing servers ($P_0$, $P_1$), they do not contain any meaningful information about the original data but random values. Meanwhile, $P_2$'s only task is to generate Beaver triples, and it receives nothing from other parties. The security proofs of computation on secret shares under the semi-honest setting can be found in \cite{knott2021}. Concerning $P_3$, although this party has access to the candidate expressions and their associated fitness values during the model training, it is still challenging, and in some cases not feasible, to retrieve the actual data, especially when the data is extensive in size and dimension. The main reason behind this challenge is that there could be numerous or even an infinite number of datasets that can generate the same fitness value for a specific expression.

\section{Preliminary experiment}

In this section, we conduct experiments on simulated data to evaluate the capabilities of the proposed framework.

\subsection{General Settings}

We evaluated GPSR and PPSR on the Nguyen benchmark suite \cite{uy2011}, a standard collection of expressions utilized in symbolic regression.  Due to the need to simulate a vertical setting in which variables are divided among different clients, we could only use benchmarks that involve multiple variables. \textbf{Table \ref{table:datasets}} provides details on the selected benchmarks. The input variables are denoted by $x_1$ and $x_2$; each contains 20 random points uniformly sampled between 0 and 1. The ground-truth expression is used to compute the target output $y$ from the input variables. The training set and test set are generated using different random seeds. While the training data is utilized to compute the fitness for each candidate expression, the test data is used at the end of the model training process to assess the best optimal expression that has been discovered.

\begin{table}[!ht]
\centering
\begin{tabular}{||c c||} 
 \hline
 Benchmark & Expression\\
 \hline\hline
 Nguyen-9 & $sin(x_1) + sin(x_2^2)$\\
 Nguyen-10 & $2sin(x_1)cos(x_2)$\\
 Nguyen-12 & $x_1^4 - x_1^3 + \frac{1}{2}x_2^2 - x_2$\\
 Friedman-2 & $10sin(\pi x_1x_2) + 20(x_3 - 0.5)^2 + 10x_4 + 5x_5$ \\
 \hline
\end{tabular}
\caption{The selected benchmark suite.}
\label{table:datasets}
\end{table}

The objective of the experiment is to compare the performance of PPSR with that of the standard GPSR. To do that, we need to simulate a vertical scenario where different clients possess different variables and then split data accordingly. As our benchmark datasets only have two input variables, we assume there are two clients, $C_1$ and $C_2$. While $C_1$ has $x_1$, $C_2$ owns $x_2$ and $y$. For each benchmark, the generated dataset is vertically divided into two hypothetical private datasets based on the variable split. Only when training GPSR models these local datasets will be re-merged to construct the centralized dataset. The same hyperparameter settings are used for both methods and are described in Table \ref{table:hyperparameters}.  

\begin{table}[!ht]
\centering
\begin{tabular}{||c c||} 
 \hline
 Parameter & Value\\
 \hline\hline
 Population size & 1000\\
 Fitness function & MSE\\
 Initialization method & Full\\
 Selection type & Tournament\\
 Tournament size & 5\\
 Mutation probability & 0.25\\
 Crossover probability & 0.95\\
 Minimum subtree depth & 0\\
 Maximum subtree depth & 2\\
 Max depth & 15\\
 Max length & 100\\
 Non-terminals & $+, -, \times, sin, cos$\\
 Terminals & $x_1, x_2, constants$\\
 \hline
\end{tabular}
\caption{Hyperparameters for the algorithms.}
\label{table:hyperparameters}
\end{table}

For each problem set, the experiment is repeated 100 times. We quantify the accuracy using $MSE$ and $R^2$ measured on both the training and test sets. In addition, the recovery rate achieved by each method will also be reported. This rate provides a more faithful evaluation of an SR method's ability to uncover the data generation process compared to the accuracy metrics \cite{la2021}. We use the most stringent recovery criteria to compute the recovery rate, i.e., exact symbolic equivalence. The equivalence of a model and the ground-truth expression is determined using a computer algebra system called SymPy \cite{meurer2017}. 

The implementation is written in Python and utilizes two libraries, CRYPTEN \cite{knott2021}, and DEAP \cite{fortin2012}. While CRYPTEN supports MPC operations, DEAP provides the basic structure for building GPSR.

\subsection{Results}

\begin{figure*}[!ht]
    \centering
    \includegraphics[width=0.83\textwidth]{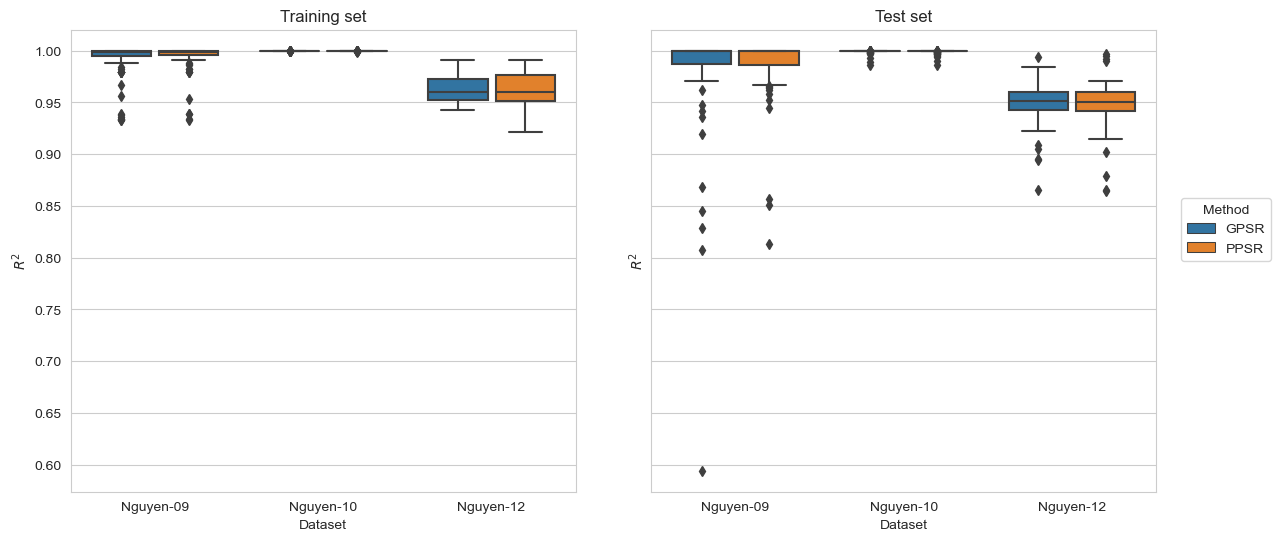}
    \caption{$R^2$ scores attained by GPSR and PPSR on the training set (left) and the test set (right).}
    \label{fig:r2}
\end{figure*}

\begin{figure*}[!ht]
    \centering
    \includegraphics[width=0.83\textwidth]{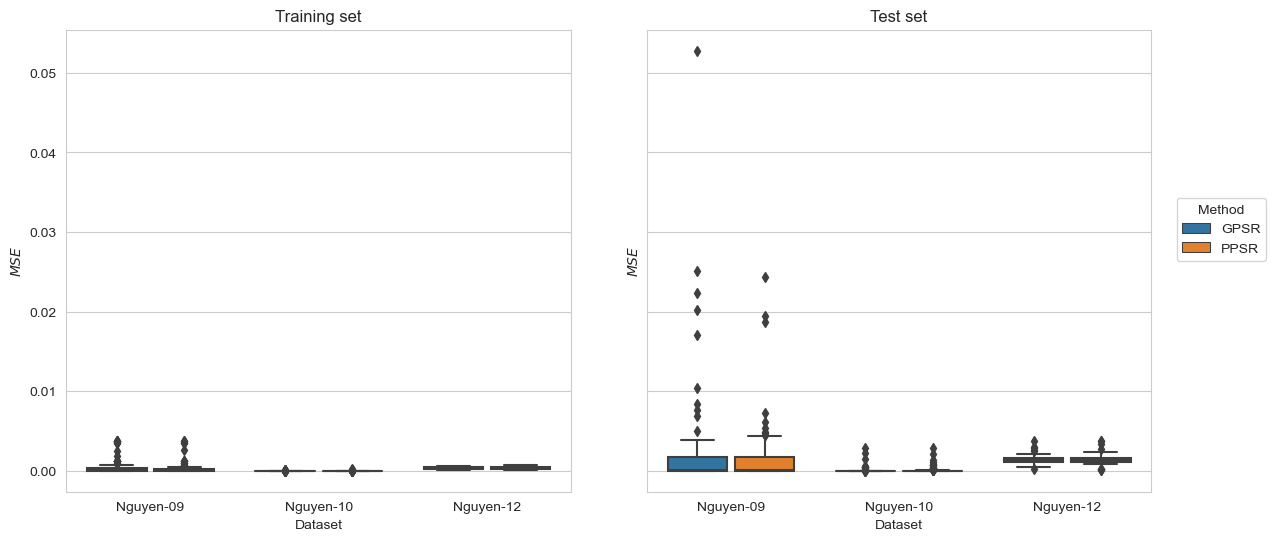}
    \caption{$MSE$ attained by GPSR and PPSR on the training set (left) and the test set (right).}
    \label{fig:mse}
\end{figure*}

The $R^2$ and $MSE$ achieved by GPSR and PPSR are plotted in \textbf{Figure \ref{fig:r2}} and \textbf{Figure \ref{fig:mse}}, respectively. Within each of these figures, there are two subplots showing the models' performance on the training set and the test set. Meanwhile \textbf{Table \ref{table:recovery}} reports the recovery rate for each benchmark. Overall, the results demonstrate that both algorithms performed effectively on the three selected benchmarks, with particular success on the Nguyen-10.

For the Nguyen-9, out of 100 runs, GPSR was able to discover the ground-truth expression 34 times, while PPSR succeeded 31 times. The top 3 most frequently found formulas for both algorithms are also very similar (see \textbf{Table \ref{table:nguyen_9_frequency}}). The Nguyen-10 problem is solved almost perfectly by both methods. For this problem, PPSR's recovery rate is 59\%, while GPSR's figure is even higher at 68\%. For the Nguyen-12, although the algorithms could not find the exact data-generating function, they could still produce highly accurate estimates for the range in which the input features are generated.

\begin{table}[!ht]
\centering
\begin{tabular}{||c c c||} 
 \hline
 Benchmark & GPSR & PPSR\\
 \hline\hline
 Nguyen-9 & 34\% & 31\%\\
 Nguyen-10 & 68\% & 59\%\\
 Nguyen-12 & 0\% & 0\%\\
 Friedman-2 & 0\% & 0\%\\
 \hline
\end{tabular}
\caption{Recovery rate comparison of GPSR and PPSR on the benchmark datasets.}
\label{table:recovery}
\end{table}

\begin{table}[!ht]
\centering
\begin{tabular}{||c c c||} 
 \hline
 Benchmark & GPSR & PPSR\\
 \hline\hline
 $sin(x_1) + sin(x_2^2)$ & 34 & 31\\
 $sin(x_1) + x_2^2$ & 7 & 3\\
 $sin(x_1) + x_2sin(x_2)$ & 4 & 5\\
 \hline
\end{tabular}
\caption{Top 3 most frequently discovered expressions on the Nguyen-9 benchmark.}
\label{table:nguyen_9_frequency}
\end{table}

Even though the solutions that GPSR and PPSR yielded are not always the same, their overall performance on the Nguyen benchmark suit is highly comparable. The advantage of using PPSR over GPSR is that our proposed approach does not require clients to expose their raw data to any entities, thus safeguarding their privacy.

\subsection{Discussions}

As PPSR relies on MPC to compute fitness values, it confronts inherent difficulties associated with this cryptographic technique. Even though MPC protects the data during computation, eventually, the clients get to view the fittest expression. Under some conditions, it might be possible for a client to infer the data of others. Take the Nguyen-9 for example, if the discovered expression is $y = x_1 + sin(x_2)$, $C_2$ can easily estimate $x_1$ since it already knows $x_2$ and $y$, i.e., $x_1 = y - sin(x_2)$. A potential mitigation for this problem is to combine MPC with Differential Privacy (DP)
\cite{pettai2015}. As DP systematically injects noise into private data before sharing, the best a malicious party can infer is the noisy values. However, implementing DP is not trivial and might come at the cost of performance.

Additionally, the current version of PPSR has to deal with numerical problems which are frequently encountered in CRYPTEN-based machine-learning implementations \cite{knott2021}. This is due in part to the use of fixed-point representation, which is more vulnerable to overflow or underflow than floating-point representations. Additionally, rounding errors can accumulate during computation and lead to inaccurate results. The debugging process for these errors can be challenging as they may only occur in the multi-party setting, in which no individual party has the ability to identify them. Therefore, implementing some commonly used mathematical operations in SR, such as $division$, $logarithm$, or $exponential$, can be difficult and requires careful consideration. 

\section{Conclusion and Future Work}

In this paper, we propose a framework for training Symbolic Regression on vertically distributed data in a privacy-preserving manner. In the proposed framework, MPC is leveraged to evaluate the quality of candidate expressions without requiring participating clients to expose their private data. The preliminary experiments on simulated data indicate that our proposed method could yield comparable performance to the centralized solution.

Although our focus in this work is the vertical setting, the proposed framework can also handle the horizontal setting. It is because PPSR relies on the 3PC scheme to conduct secure fitness computation, and 3PC has no assumptions about how data is distributed. Besides, since there are no constraints about the evolution operators, our method can be easily extensible to other GPSR variations. For future work, we will explore these extensions. Furthermore, we also plan to investigate the combination of MPC and DP to strengthen the safety of the framework and to examine potential solutions for the numerical issues to increase precision. Finally, we will conduct experiments on more complex benchmarks that better reflect reality to obtain a more comprehensive evaluation of the capability of the proposed method.


\bibliographystyle{ACM-Reference-Format}
\bibliography{sample-base}

\appendix

\end{document}